\documentclass[11pt,twoside]{article}

\usepackage{asp2006}
\usepackage{amssymb}
\usepackage{psfig}
\usepackage{epsfig}
\usepackage{graphicx}
\usepackage{lscape}

\begin{document}

\markboth{Moran}{Accretion Disk in NGC\,4258}
\pagestyle{myheadings}
\setcounter{equation}{0}
\setcounter{figure}{0}
\setcounter{footnote}{0}
\setcounter{section}{0}
\setcounter{table}{0}

\title{The Black Hole Accretion Disk in NGC\,4258: One of Nature's Most Beautiful Dynamical Systems}
\author{James M. Moran} 

\affil{Harvard-Smithsonian Center for Astrophysics, USA}  

\begin{abstract} 

In this talk I will summarize some of the work that the CfA group has done to study the structure of the water masers in the accretion disk of NGC\,4258. A series of 18 epochs of VLBA data taken from 1997.3 to 2000.8 were used for this study. The vertical distribution of maser features in the systemic group was found to have a Gaussian distribution, as expected for hydrostatic equilibrium, with a $\sigma$-width of 5.1 $\mu$as. If the disk is in hydrostatic equilibrium, its temperature is about 600K. The systemic features exhibit a small, but persistent, gradient in acceleration versus impact parameter. This characteristic may indicate the presence of a spiral density wave rotating at sub-Keplerian speed.  A more precise understanding of the dynamical properties of the disk is expected to lead to a more refined estimate of the distance to the
galaxy.
\end{abstract}

Please forgive my slightly flamboyant title, it's really not me, but it reveals a little bit about how enamored I have been with the galaxy NGC\,4258 for almost 15 years now since the Nobeyama group \citep {moran:nakai93} first discovered the high velocity maser features in its water vapor spectrum, which are  symmetrically placed with respect to the systemic masers and which provide evidence of a rotating disk structure.  VLBA measurements were able to confirm that proposal some years later.

There are several reasons why we want to keep hammering away at NGC 4258. Firstly, it is a very beautiful, simple system and there is a lot of physics that we can study in the inner 1~parsec of its accretion disk around its central 
black hole.  Secondly, as Jim Braatz showed in his talk
earlier today \citep{moran:braatz08}, this is the prototype source, still the flagship example,
of a simple, clean megamaser system in a galaxy, so it is a laboratory in which we can study all
\begin{figure}[!ht]
\begin{center}
\includegraphics[width=12cm]{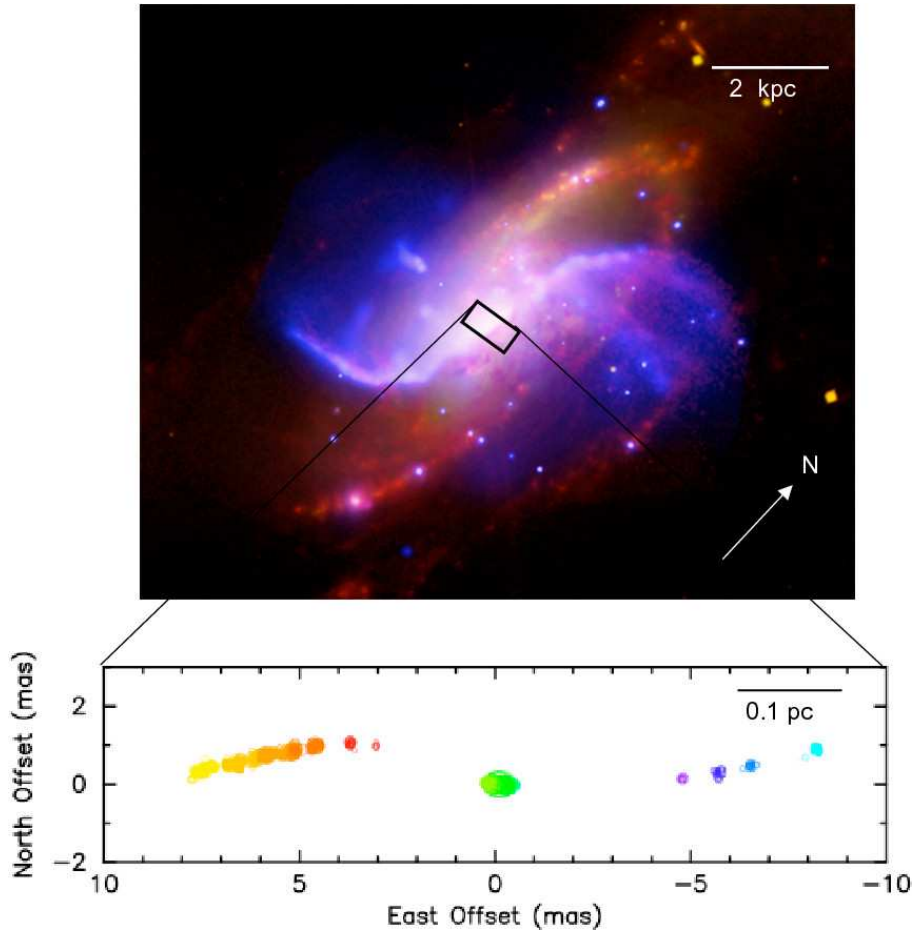}
\caption{Top: A multiwavelength image (X-ray, H$\alpha$, 1.4 GHz radio) from \citet{moran:yang07}
of the inner part of the galaxy NGC\,4258, also known as M106. Unfortunately, this black-and-white rendition of the color image does not serve to distinguish the various wavelength regimes very well. The so-called anomalous arms, seen by their synchrotron emission, have position angles of about $-$45 and
135 degrees and exhibit sharp bends about 2 kpc from the nucleus. A large angle between the
position angle of the accretion disk and the galaxy was surmised by \citet{moran:oort82}.
Bottom: Image of the maser emission from the central accretion disk of the galaxy from VLBA
observations. Note that the overplotting of many maser emission spots makes the disk look much
fatter than it actually is. From \citet{moran:argon07}
\label{jmm:fig1}}
\end{center}
\end{figure}
the small effects that can affect the usefulness of megamasers as distance indicators. As Braatz mentioned, it's rather difficult to find masers like NGC\,4258: several thousands of galaxies have been surveyed and the total count of megamaser galaxies is now about a hundred. This is about a 5\% yield of galaxies searched, possibly because only masers in edge-on accretion disks can be detected. And then, of those hundred, only about 5 percent, four or five,
are probably going to be good distance indicators because their geometry can be readily understood. So the yield of galaxies suitable for precise distance measurement is actually quite small so far.

What I want to describe today is the large set of observations that has been analyzed by
our group at CfA -- Alice Argon, Lincoln Greenhill, Elizabeth Humphreys, myself and Mark Reid. I will also talk about some of the Ph.D. projects of Harvard students over the past decade -- Jim
Herrnstein originally measured the distance precisely to NGC\,4258 and fully characterized the warp of the accretion disk, Adam Trotter refined the warp model, Ann Bragg did an analysis of the maser accelerations with the VLA, and Maryam Modjaz put an upper limit on the Zeeman effect from circular polarization measurements with the GBT, 
which yielded a limit on the magnetic field strength.

Figure~\ref{jmm:fig1} shows a multiwavelength composite image of NGC\,4258. In the optical part of
the spectrum the galaxy looks like a normal massive spiral galaxy with an indication of a central
bar. However, its strong nuclear emission lines led \citet{moran:seyfert43} to include it in his
original catalog of disturbed galaxies, which became known as Seyfert galaxies. The radio
continuum emission, first imaged in high resolution at Westerbork at 1.4 GHz by \citet{moran:vanderkruit72}, shows a pair of anomalous arms of synchrotron emission, which so intrigued
Jan Oort many years ago. In one of the last papers of his career \citep{moran:oort82} he
reminisced very insightfully: ``One reason why the peculiar phenomena [anomalous arms] that are so striking in NGC\,4258 do not occur more often in spirals may be that they can occur only when the axis of
ejection (presumably the rotation axis of the central black hole's accretion disk) is almost
perpendicular to the rotation axis of the galaxy.'' As we determined in 1995, the angle between
the axis of the accretion disk, as defined by the masers, and the rotation axis of the galaxy is
about 120$\deg$ \citep{moran:miyoshi95}.

\begin{figure}[ht]
\begin{center}
\includegraphics[width=12cm]{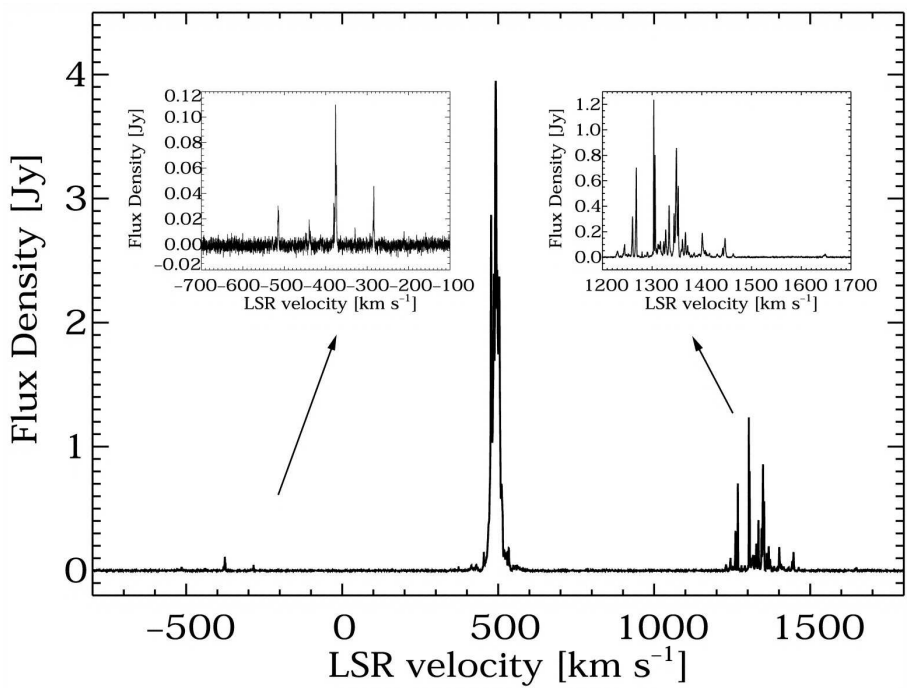}
\caption{A spectrum of the water vapor emission from NGC\,4258 made with the GBT. Velocity axis (LSR, radio definition) is measured with respect to the rest frequency of 22235.080 MHz. The insets show blowups of the high velocity red- and blueshifted portions of the spectrum.
\label{jmm:fig2a}}
\end{center}
\end{figure}

\begin{figure}[ht]
\begin{center}
\includegraphics[width=5cm]{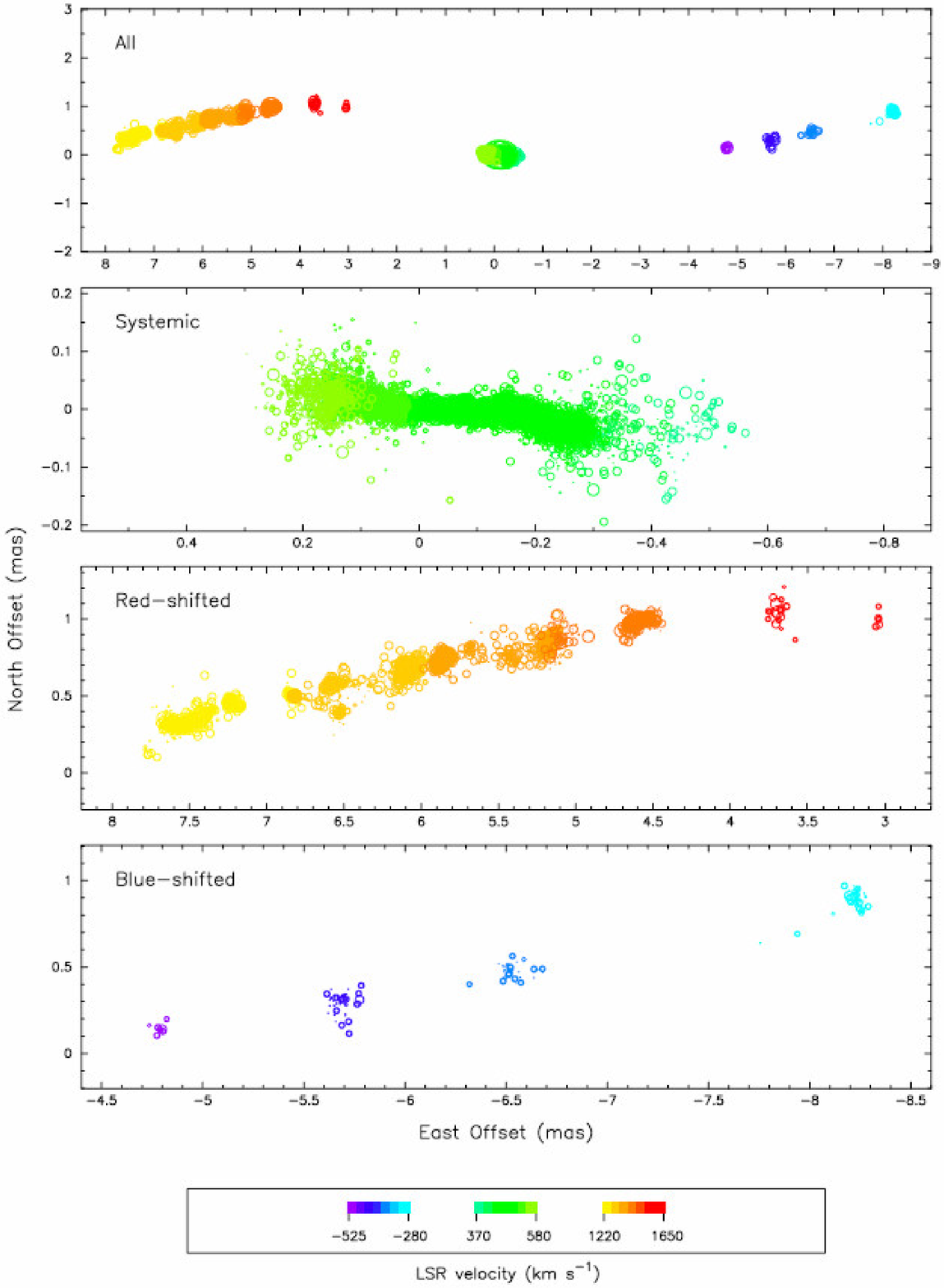}
\hfill
\includegraphics[width=7cm]{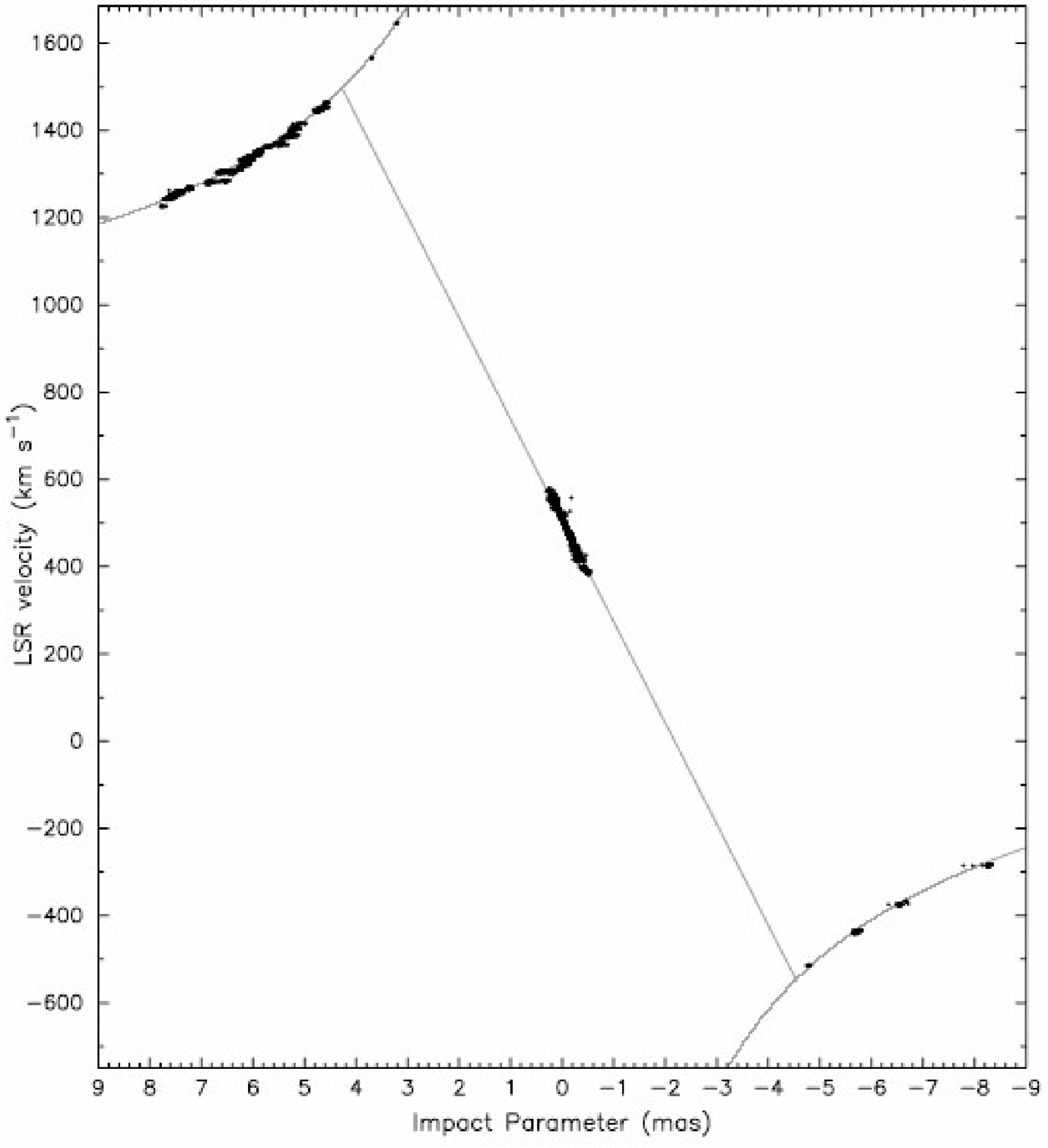}
\caption{Left: Images of maser emission from NGC\,4258 averaged over 18 epochs, 1997.3 -- 2000.8. 
Right: Position-velocity diagram of the masers along the axis of disk at a position angle of 85 degrees. Note the features near 1600 km s$^{-1}$ that lie inside the radius of the systemic features. From \citet{moran:argon07}.
\label{jmm:fig3}}
\end{center}
\end{figure}

Figure~\ref{jmm:fig2a} shows a spectrum of our archetypical maser.  You see the nicely symmetric
bands of features around the galaxy's systemic velocity of about 
500 km s$^{-1}$: the high redshifted
velocity features near 1500 km s$^{-1}$ and the high blueshifted velocity features near $-500$~km~s$^{-1}$.   \citet{moran:nakai93} originally suggested three possible explanations for the origin of this spectral structure: (1) a rotating disk, (2) a bipolar outflow, and (3) a non-kinematic process--stimulated Raman emission. The first direct evidence that rotation was the correct interpretation came from the VLBI observation of the systemic features by \citet{moran:greenhill95} from data taken in 1984.  One of the first projects executed by the 
VLBA showed the full distribution of the masers, which clearly defined a disk in Keplerian motion and really captured the imagination of the astronomical community \citep{moran:miyoshi95}. As shown in Figure~\ref{jmm:fig1},
the maser distribution follows a perfect monotonic progression in velocity for each of the three velocity segments indicative of a rotating, slightly warped, Keplerian disk whose diameter is about 0.5 pc. Figure~\ref{jmm:fig3}, shows an expanded version of the image and the position-velocity (p-v) plot of the masers, which reveals the Keplerian signature.

The VLBA, a wonderfully precise astrometric machine, which works so reliably and has a resolution of 200~$\mu$as at the water maser frequency of 22 GHz, made these measurements possible.  Another important feature of the VLBA is that  because it has a very large correlator its spectral resolution is very fine, better than 1 km s$^{-1}$, which allows  resolution of all of the maser spectral features and complete coverage of the spectrum. From 1997 to 2000 we embarked on a very ambitious set of 18 VLBI experiments and have just now gotten all the data published \citep{moran:argon07}. Some 14,000 maser features have been detected: 10,039 systemics, 3,979 reds and about 289 blues.  There are very few blue features (see Fig.~\ref{jmm:fig1}) but now we have enough to solve accurately for a kinematic model. These 14,000 features (or velocity channels) are not all distinct maser spots because there are several velocity channels across each line profile and the features show up in multiple epochs.  There are probably about a thousand independent maser spots that we have mapped among these epochs.  All this data is now online in the paper by \citet{moran:argon07} and I
invite anyone who would like to look for interesting effects in NGC\,4258 to use the archive. Please, feel free to dig in!

\begin{figure}[ht]
\begin{center}
\includegraphics[width=12cm]{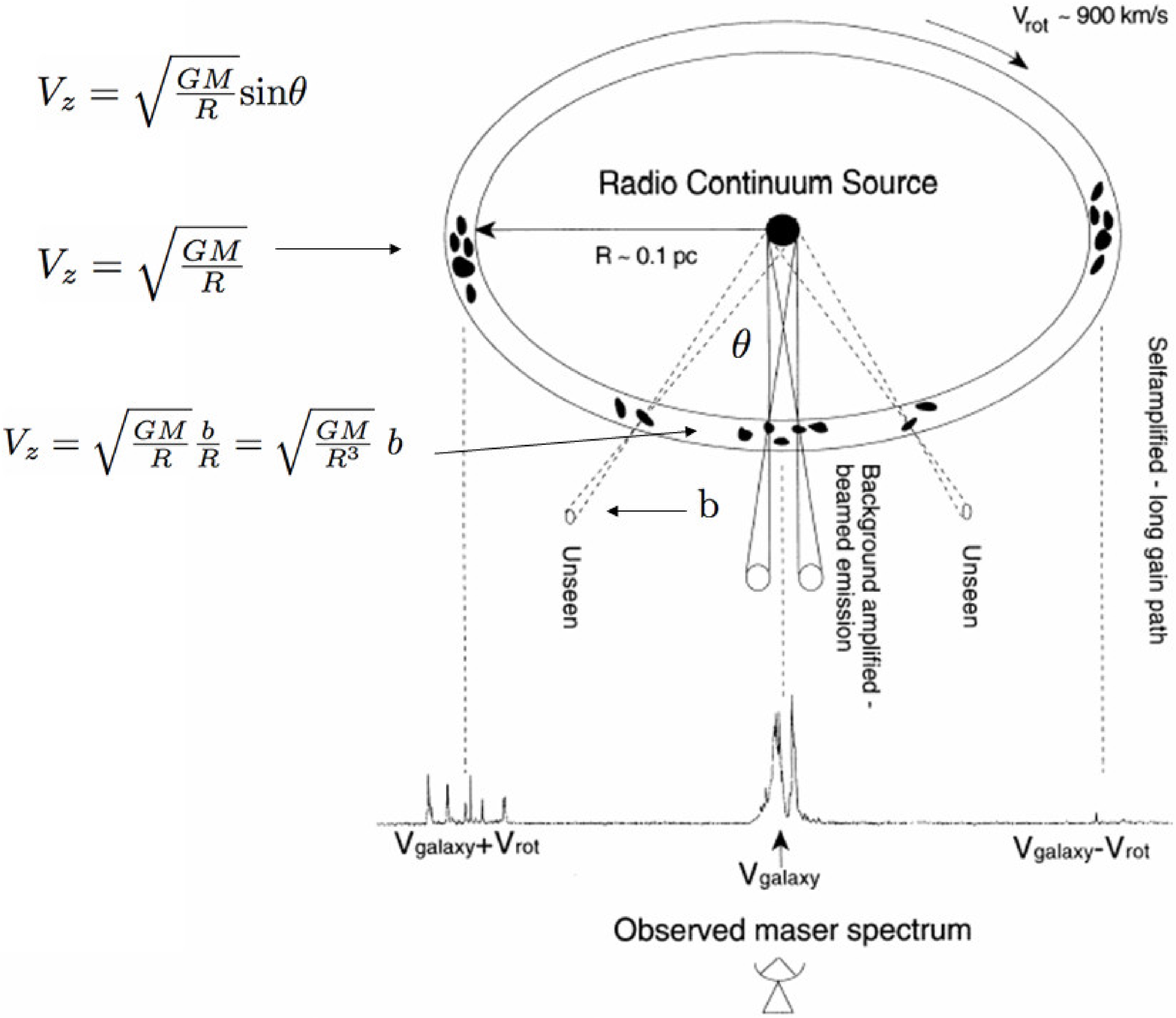}
\caption{Cartoon of the maser disk in NGC\,4258. Adapted from \citet{moran:greenhill95}.
\label{jmm:fig4}}
\end{center}
\end{figure}

Figure~\ref{jmm:fig3} is an ensemble of all the data showing the disk over half of a parsec in diameter in the top panel on the left.  The lower panels are blow-ups around the three velocity ranges. Note the narrow waist in the systemic velocity range. The important thing is that the position-velocity diagram in Figure~\ref{jmm:fig3} (right) shows a nearly perfect Keplerian disk in the high velocities. The linear part in the center shows that the radial distance of those masers from the nucleus is
nearly constant. Figure~\ref{jmm:fig4} is a cartoon that we have used for many years to explain the 
position-velocity characteristics of the masers. The line-of-sight velocity, $V_{\rm z}$, in a Keplerian disk viewed edge-on is given by $V_{\rm z} = \sqrt{GM/R} \sin \theta$ where $\theta$ is the azimuth angle in the disk measured from the line-of-sight to the 
black hole, $R$ is the radial distance from the black hole of mass $M$, and $G$ is the gravitational constant.  The high-velocity features show a nearly perfect Keplerian curve $V_{\textrm z} \sim \sqrt{GM/R}$ and therefore they must be at a fixed azimuth angle in the disk. Since the accelerations of these features are small, the azimuth angle is probably close to 90$\deg$, i.e., these features are on the so-called `mid-line' where $\sin\theta = 1$. Close to the line of sight in front of the black hole $\sin\theta = b/R$ where $b$ is the impact parameter so $V_{\textrm z} = b\sqrt{GM/R^3}$.  Hence there is a linear relationship between the line-of-sight velocity and the impact parameter $b$ for the case where the radius $R$ is fixed. This describes the p-v diagram systemic masers very well (see Fig.~\ref{jmm:fig3}).

It is interesting to note that we observe masers in front of the black hole and at the $\theta = \pm90\deg$ azimuth angles. The velocity gradient is zero in these directions and that is undoubtedly an important factor in why we see masers in just those directions.  We have always been vigilant about looking for masers on the back side of the disk. They would be slightly above the ones in front because of the disk inclination, and would be accelerating in the opposite direction. We have never identified any back-side features.  

Inspection of spectra from 1982 onward show that the systemic masers are always confined to a velocity range of about 430 to 530 km s$^{-1}$. They are like toy soldiers, they first appear near 430 km s$^{-1}$, march uniformly in velocity at about 8 km s$^{-1}$ yr$^{-1}$ until they reach the upper velocity limit of about 530~km~s$^{-1}$ in about 12 years, after which they disappear. We infer from this behavior that the maser emission is beamed
towards us with a beam angle of about 8$\deg$, which causes them to be visible to us only in a relatively narrow velocity range corresponding to $| \theta | < 4^{\circ}$.  If it weren't for turbulence, we might see the currently visible masers again in about a thousand years when they rotate back into view.  All of the masers that we mapped originally in 1995 are gone from our view, and presumably pointing their radiation elsewhere in the universe.

\begin{figure}[!ht]
\begin{center}
\includegraphics[width=12cm]{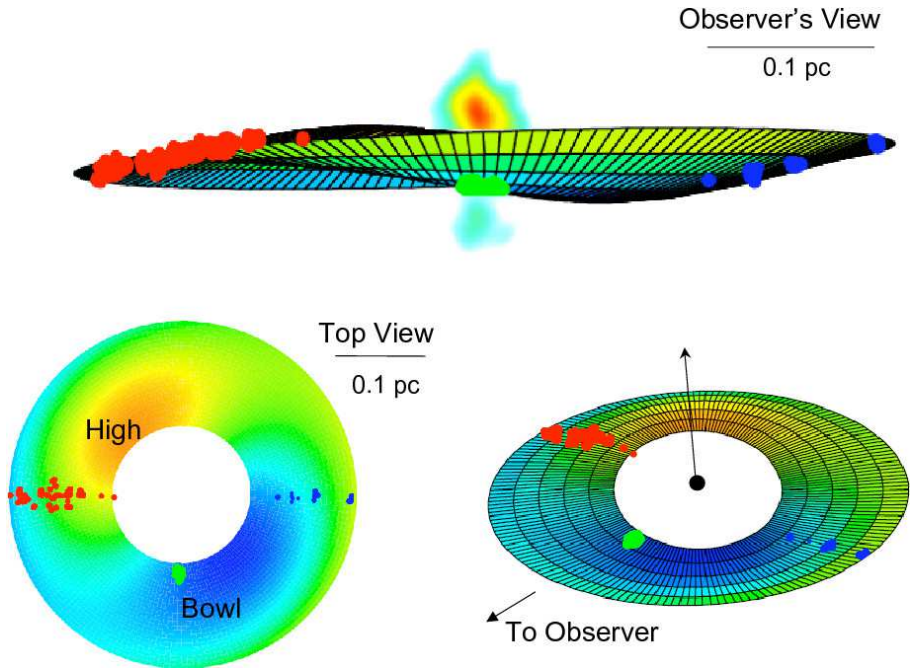}
\caption{Three views of the maser disk in NGC\,4258 with the warp parameters of 
\citet{moran:herrnstein05}.  The upper model (observer's view) shows the 1.3-cm continuum emission that is aligned along the spin axis of the disk. \label{jmm:fig5}}
\end{center}
\end{figure}

We have been able to characterize the warping of the disk fairly precisely \citep{moran:herrnstein05}. Figure~\ref{jmm:fig5} shows the disk from several vantage points.  The spread in azimuth angles of the red features is inferred because the line-of-sight components of the accelerations are not zero [i.e., $a_{\rm z} = (GM/R^2)\cos\theta$].  The warp is completely characterized by data in one plane -- the plane perpendicular to the line-of-sight encompassing the midline.  The position-angle warp is obvious in Figure~\ref{jmm:fig1}. There also seems to be a warp in inclination angle as evidenced by the very slight systematic deviations from Keplerian motion among the high velocity masers.  The motion is Keplerian to about a part in a hundred, but there is a systematic deviation in velocity totaling about 9 km s$^{-1}$, which appears as a flattening of the Keplerian curve and which we interpret as an linear change in inclination angle with radius.  The parameters of this warp are given in \citet{moran:herrnstein05}. They describe a disk that has the characteristic of having a ``bowl'' in the direction of the observer (see Fig.~\ref{jmm:fig5}), which probably explains why the systemic features are confined to a narrow annular radius. The warp of the disk could have various causes. Perhaps, most intriguingly, it could be due to interaction of the accretion disk and the spinning black hole if their spin axes are misaligned, i.e., the Bardeen-Peterson effect \citep{moran:bardeen75} as suggested by \citet{moran:moran99}. Recent quantitative calculations by \citet{moran:caproni07} and \citet{moran:martin08} show that this effect is quite viable and the critical transition radius is about 0.1 pc, the inner radius of the maser disk.

\begin{figure}[hb]
\begin{center}
\includegraphics[width=10cm]{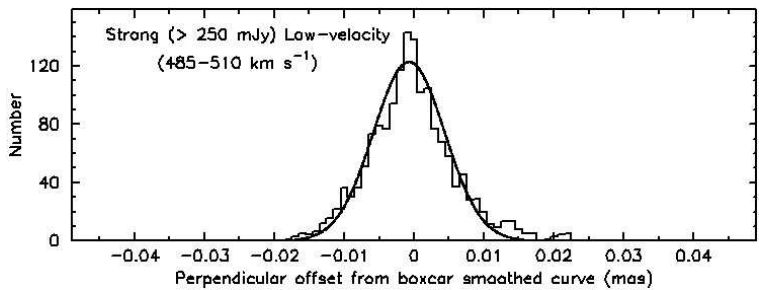}
\caption{Measurement of the thickness of the accretion disk in the vicinity of the systemic
features, showing $\sigma = 5.1~\mu$as. From \citet{moran:argon07}.\label{jmm:fig6}}
\end{center}
\end{figure} 

We have always wondered exactly how thick the disk is \citep[e.g.][]{moran:moran95}. So, to make this measurement, we have looked right at the sweet spot in the disk at the bottom of the bowl. There, the disk is viewed edge-on and the differences in radial positions in the masers don't cause a spread in projected vertical distance.  We have sorted the vertical positions of the strong
masers with respect to the disk and made a histogram of their vertical distribution.  The measurement errors in these relative positions are about one  $\mu$as, which is really quite phenomenal.  The histogram in Figure~\ref{jmm:fig6} shows that the height distribution is nearly a Gaussian function with a $\sigma$-width of about 5~$\mu$as. If this disk is in hydrostatic equilibrium, we would expect a Gaussian distribution with the width equal to R, the radius of the systemic masers at the bottom of the bowl, times the sound speed $c_{\textrm s}$ divided by the Keplerian velocity $v$, i.e., $\sigma = Rc_{\textrm s}/v$. In this case, the required sound speed is about 1.5 km s$^{-1}$ corresponding to a temperature of about 600K. The disk is about 5~$\mu$as thick at a radius of about 4 mas from the center, so it is really a remarkably thin disk, with a fractional radius of about one part in a thousand, just as simple theory suggests \citep{moran:frank02}.

\begin{figure}[ht]
\begin{center}
\includegraphics[width=12cm]{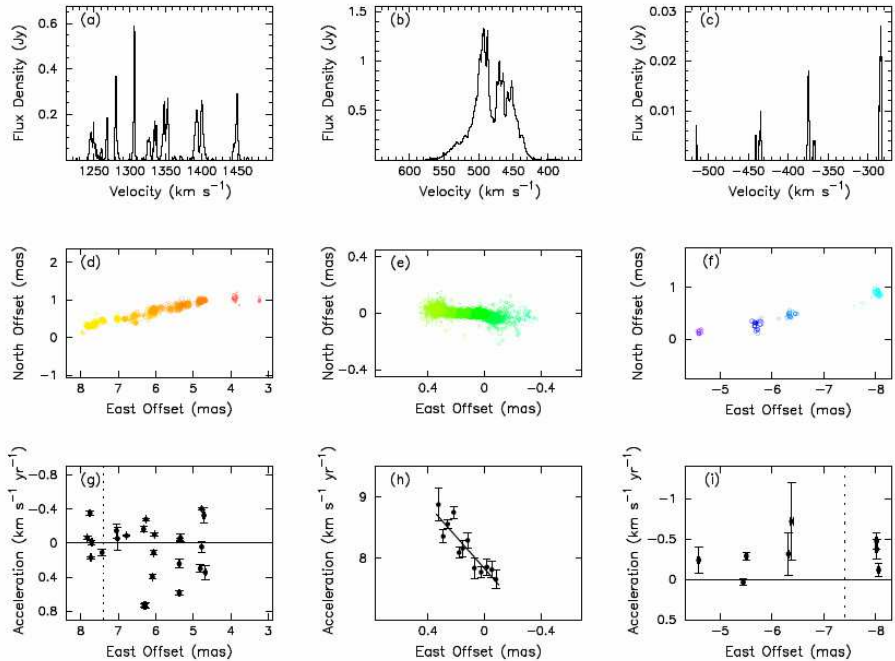}
\caption{Velocities, positions and accelerations of masers in NGC\,4258.  The vertical dotted lines at $\pm7.4$ mas in panels g and j, mark the position where the disk is viewed exactly edge-on according to the warp model. From \citet{moran:humphreys08}.
\label{jmm:fig7}}
\end{center}
\end{figure}

\begin{figure}[htp]
\begin{center}
\includegraphics[width=4cm,angle=-90]{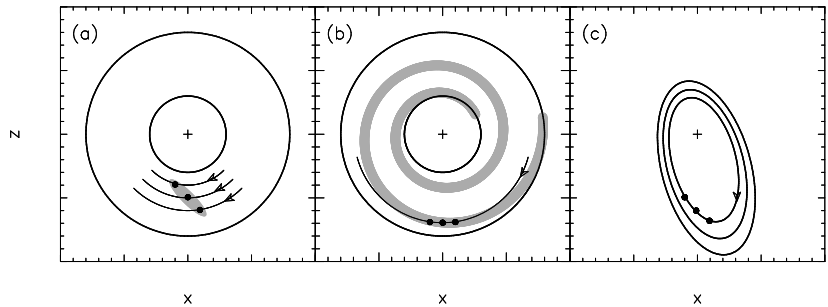}
\caption{Three proposed explanations for the persistence of the acceleration trend among the systemic features shown in Figure~\ref{jmm:fig7} (panel h). (a) The gray patch is a stationary preferred location for maser
emission. (b) Spiral density wave in gray, which gives extra acceleration to approaching masers. (c) Elliptical orbits. From \citet{moran:humphreys08}. \label{jmm:3panel}}
\end{center}
\end{figure} 

We have also tried to measure the magnetic field in the masers with both the 
VLA and the 
GBT via the
Zeeman effect \citep{moran:modjaz05}.  H$_{2}$O is a non-paramagnetic molecule so the Zeeman effect
is not very sensitive to the magnetic field. We determined a limit on the toroidal component of the magnetic field of about 30 milligauss. The total magnetic field is probably less than about 100 milligauss.  This means that the total magnetic support of the disk is probably less than the thermal support but it would be interesting to pursue this a bit further and actually detect the Zeeman effect.  For a magnetic field of about 100 milligauss and with the assumption of equipartition of energy, a very stringent limit can be set  on the mass accretion rate of about $10^{-4}$ solar masses per year, which is interestingly low.

\begin{figure}[ht]
\begin{center}
\includegraphics[width=12cm]{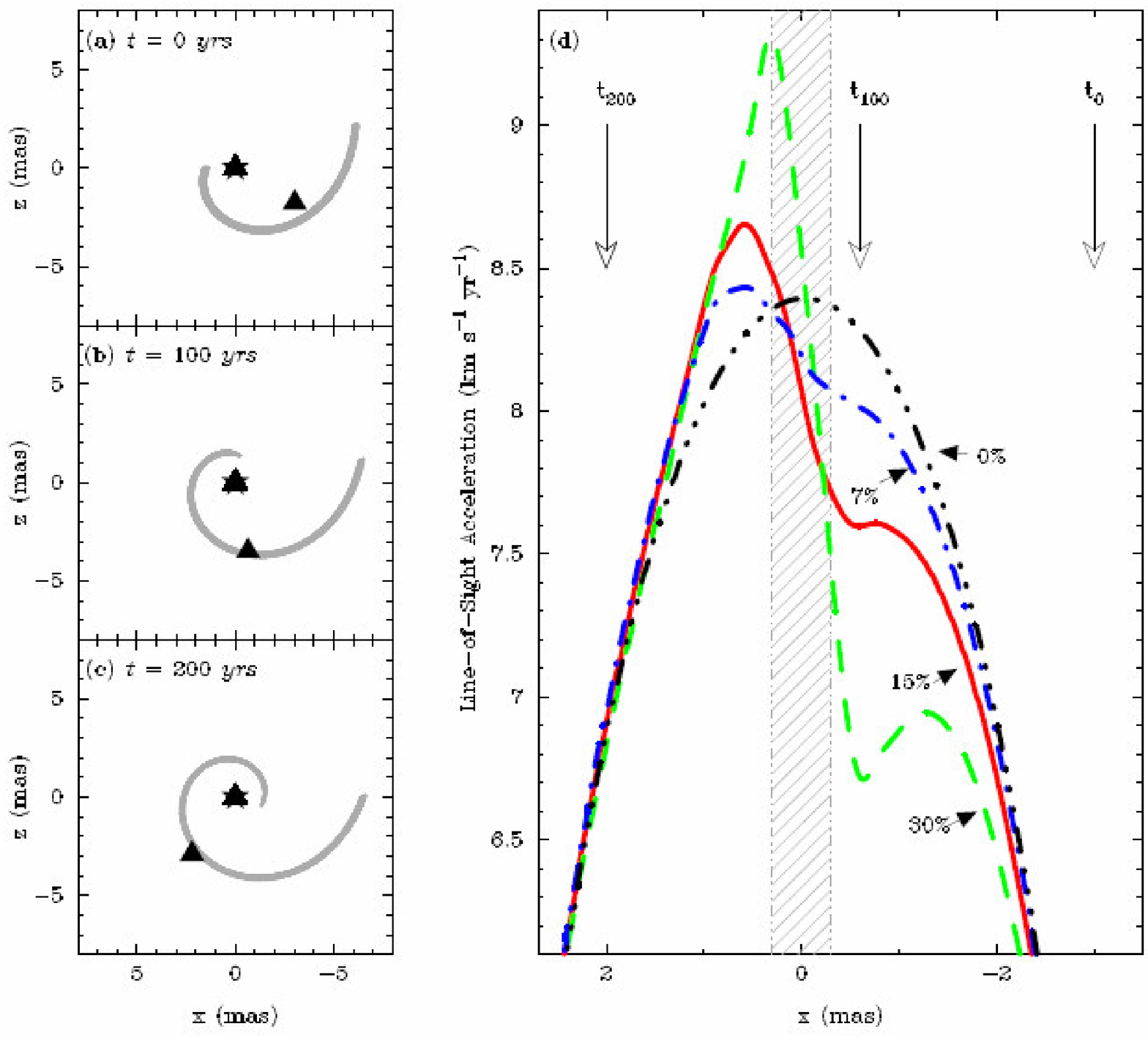}
\caption{The effect of a spiral arm on the velocities of systemic maser features. Left: time sequence of a maser ($\blacktriangle$) moving through a spiral arm.  Right: acceleration profiles for spiral arms of a given fraction of disk mass. From \citet{moran:humphreys08}. \label{jmm:fig9}}
\end{center}
\end{figure} 

Getting back to looking at the structure of the disk, Figure~\ref{jmm:fig7} shows the velocities, positions and accelerations of all of the features we have measured over the 18 epochs. The ``Watson notch'' \citep{moran:watson94} has also persisted over all these years: there is always a dip near 472 km s$^{-1}$ (see Fig.~\ref{jmm:fig7}, panel b) which some of us think may be an enduring characteristic of the systemic group.  They hypothesized that there is an absorbing layer of water farther out in the disk that absorbs at the rest frequency of the galaxy. Maoz originally proposed in 1995 \citep{moran:maoz95} that there seemed to be a spatial periodicity among the high velocity features.  We were originally skeptical of this interpretation, but the periodicity has persisted and is more evident than ever (see Fig.~\ref{jmm:fig7}, panels a, d, c and f). There seems to be a missing feature in the gap at 330 km s$^{-1}$ (see Fig.~\ref{jmm:fig7}, panel f), but we took a
very sensitive spectrum  with the GBT 
recently and found some masers right in this spot! So the
spatial periodicity seems to be real and there probably is some spiral structure in this disk. 

It is interesting that the systemic masers drift at 
8 km s$^{-1}$ yr$^{-1}$ in the mean but there is actually a slight variation across the face of the disk (see Fig.~\ref{jmm:fig7}, panel h).  This is perhaps a little bit unsettling.  The simplest explanation for this change in acceleration is that the features actually lie at slightly different radii, so you can map acceleration into radius by the relation $a_{\textrm z} = (GM/R^{2})\cos\theta$ so a 10\% variation in the acceleration gives about a 5 percent variation in the annular radial distance of the masers.  We looked at some of the old published data \citep[see][]{moran:humphreys08} and this characteristic has been noticeable since the 1980s. It seems to be a stationary pattern that persists in the disk and does not move with the rotation of the galaxy.  Figure~\ref{jmm:3panel} shows three possible explanations for this persistent acceleration slope. The simplest explanation, depicted in panel (a), is that the disk has a structure, e.g., a bump or tangent locus, that is stationary or rotates much slower than the Keplerian rotation.  We see masers only from gas that flows through this special region. Alternatively, (panel b) there may be massive spiral arm which causes a real acceleration effect as the maser blobs move through it.  So for example in Figure~\ref{jmm:fig9}, if the ``star'' symbol is the black hole and the curved arc is the spiral arm moving at half the rotation speed, a maser (triangle) coming onto the spiral arm will be accelerated towards that spiral arm with an extra acceleration in addition to the Keplerian acceleration. The right panel shows accelerations for various values of disk mass. For seven percent of the maximum disk mass that's allowed by the Keplerian curve (about 1\% of the black hole mass), the acceleration trend across the observable range of $\pm0.4$ mas matches the observations.  Over a long time this acceleration profile would shift and allow the pattern speed of the spiral arm to be determined. A third explanation (Fig.~\ref{jmm:3panel}, panel c) for the acceleration anomaly is that the orbits are slightly elliptical.

The high velocity features have accelerations that have been very accurately measured.  The features are all separated in velocity and we can measure the drift in velocity with time very accurately.  Each of the accelerations are precisely known but are scattered rather randomly (see Fig.~\ref{jmm:fig7}, panels g and j).  We interpret this as the features being ahead or behind of the mid-plane exactly at 90$\deg$ to us,
where they would have zero accelerations.  They have accelerations of about 0.5~km~s$^{-1}$~yr$^{-1}$ with respect to the systemic value of 8~km~s$^{-1}$~yr$^{-1}$ so they are a few degrees off the mid-line.  There is a theory
by \citet{moran:maoz98} suggesting that these accelerations are really due to a phase effect in the masers.  If there is a spiral arm passing through the disk, the accompanying shock wave may change the point of maser excitation along the mid-line where the velocity gradient is zero.  This will move the site of an emission spot outward and to lower absolute velocity on the Keplerian curve.  We see no evidence of this. It should have been a small effect of about 0.05 km s$^{-1}$ yr$^{-1}$. If it's there, it is completely buried in the real geometric dispersion of features.

\begin{figure}[ht]
\begin{center}
\includegraphics[width=10cm]{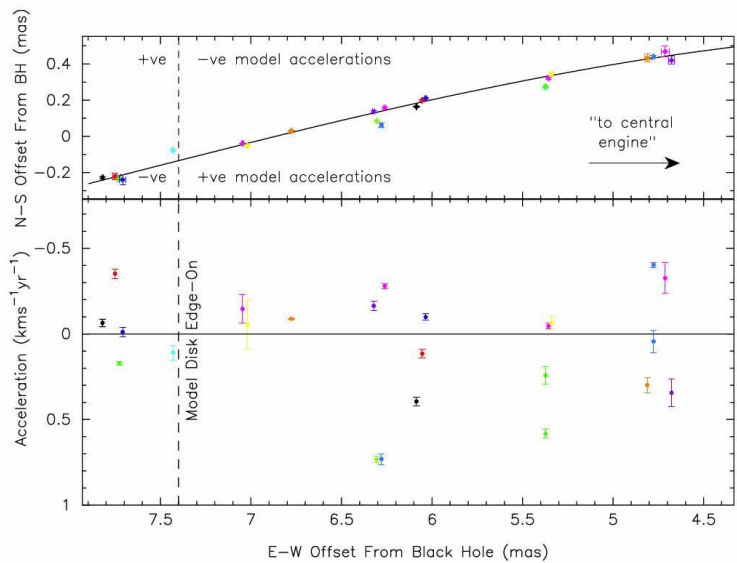}
\caption{Sky positions and accelerations in the redshifted side of the disk.  Upper panel: Sky positions of maser components as a function of East-West offset; the line indicates where the disk intersects the plane perpendicular to the line-of-sight through the black hole as predicted by the warp model of \citet{moran:herrnstein05}. Lower panel: Line-of-sight accelerations; features on the mid-line should have zero line-of-sight velocity. From \citet{moran:humphreys08}.
\label{jmm:fig10}}
\end{center}
\end{figure} 

\begin{figure}[ht]
\begin{center}
\includegraphics[width=8cm]{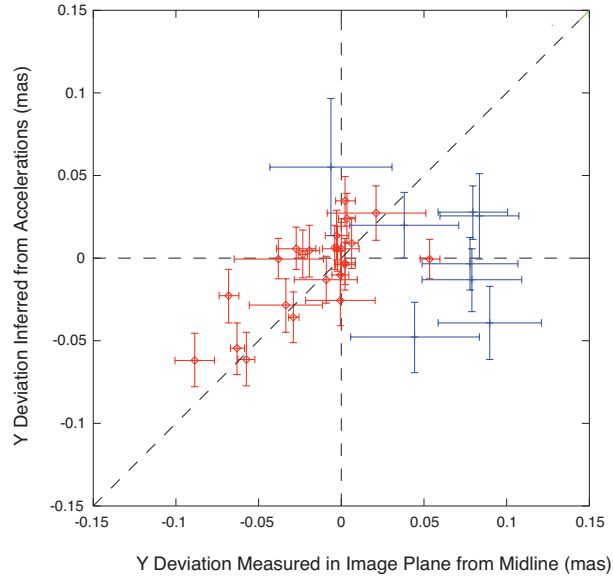}
\caption{Vertical deviations from the plane of the disk predicted by the acceleration data and warp model compared with vertical deviations measured directly from the images.  The redshifted masers are shown with data points and the blueshifted masers are shown without data points at the centers of their error bars.  The correlation, especially among the redshifted masers, provides strong support for masers being good dynamical probes of the disk.  However, the segregation of the red and blue masers suggests an unmodeled asymmetry in the disk. From
\citet{moran:humphreys08}.
\label{jmm:fig11}}
\end{center}
\end{figure} 

Figure~\ref{jmm:fig10} is a blow-up of the redshifted side of the disk. It shows the disk model fit through the maser points 4 to 8 mas from the black hole with the vertical position deviations
from the disk in the upper panel and the accelerations in the lower.  There is a high correlation between the accelerations of the features and their offsets from the disk. Figure~\ref{jmm:fig11} plots the deviations in the image versus the deviations that we would expect if we assume that the accelerations are caused by the azimuths not being exactly 90$\deg$. If our warp model is correct, then there should be a good correlation.  I draw your attention to the red features, which show a very highly correlated between these two estimates of the offset from the disk plane. The blue
features are not so well correlated because the red features were used to determine the warp characteristics. The importance of Figure~\ref{jmm:fig11} is that it shows that the geometric model of a warped disk  with masers being density enhanced blobs that physically move and can be used as dynamical tracers is largely correct. This result is reassuring for the dynamical interpretation of the maser data. 

\begin{figure}[ht]
\begin{center}
\includegraphics[width=10cm]{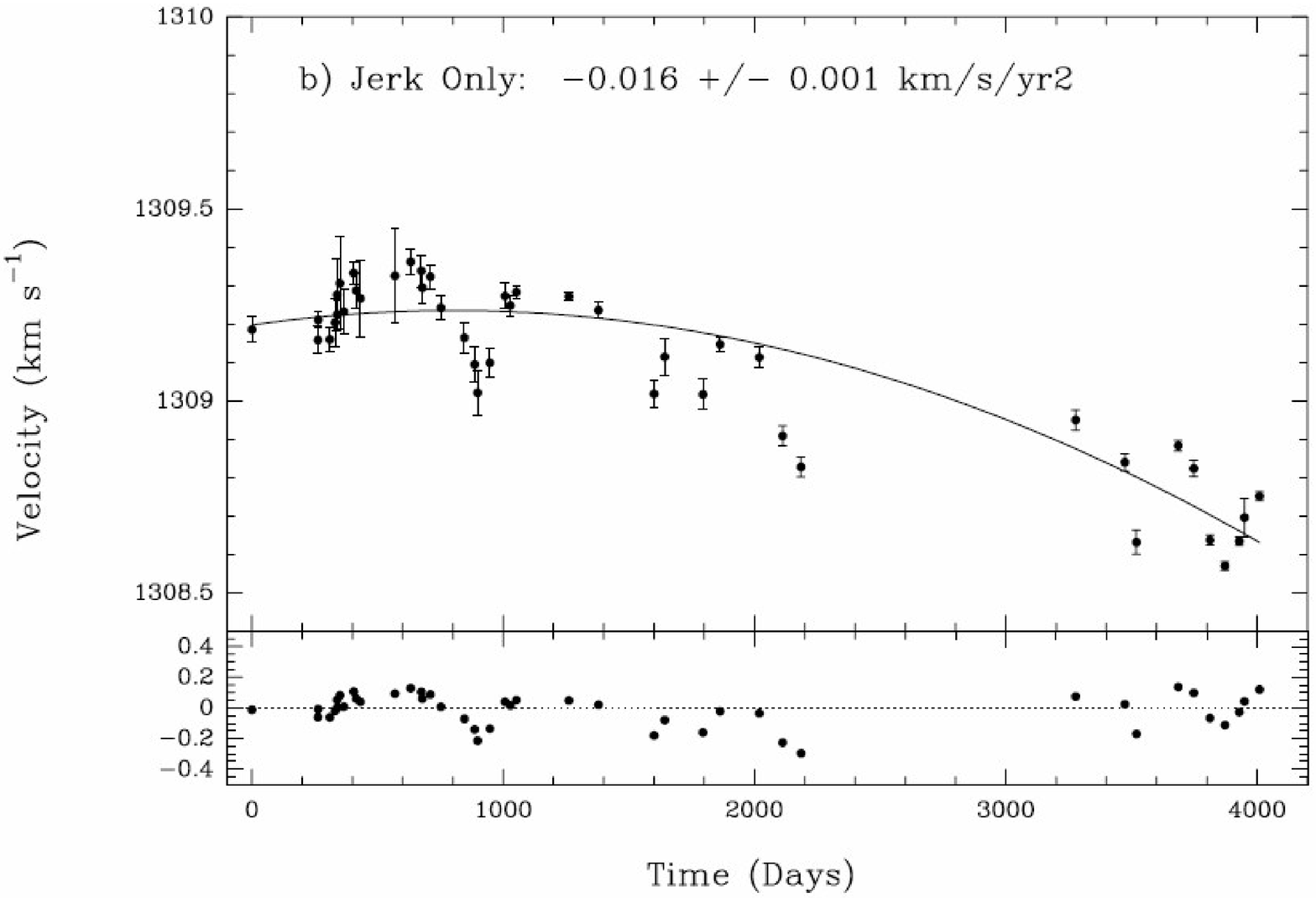}
\caption{Acceleration of the ``1309'' km s$^{-1}$ feature. The velocity is corrected for special relativistic
effects. The time axis is relative to Jan 1, 1994. Not enough data has been obtained to convincingly
demonstrate a non-linear trend in velocity. The curve shown is a fit to a sinusoid corresponding to an orbital
period of 1830 years. The bottom plot shows the residuals to the fit. The epoch of mid-plane crossing of this feature is about 1997.0. \label{jmm:fig12}}
\end{center}
\end{figure} 

And for my younger colleagues in the audience, I show in Figure~\ref{jmm:fig12} how we have tracked one particular high-velocity feature for twelve years. We have mostly been fitting straight lines to the velocity data to estimate the acceleration, but the velocity should actually follow a thousand-year sinusoid.  The line in the figure leads the eye a little deceptively, but
we may actually just be seeing a bit of that sine wave. Its maximum corresponds to the time that the maser is going through the line perpendicular to the line of sight, i.e., crossing the mid-line.  So this plot will get better quadratically with time, and can be expected to show whether the geometric model of the maser is correct.

Finally, let me bring you up to date on the distance measurements of NGC\,4258.  We announced a distance in 1999 of $7.2$ Mpc with statistical error of $\pm 0.3$ Mpc and a systematic error of $\pm 0.4$ Mpc \citep{moran:herrnstein99}. The systematic error was largely due to uncertainty in the departure of the orbits from circularity.  If the orbits are slightly elliptical, that could bias the distance.  Several groups have been looking at the Cepheid variables in NGC\,4258 with the Hubble telescope and the distance from those has been estimated as $8.1 \pm 0.4$ Mpc \citep{moran:maoz99}, $7.8 \pm 0.3$ Mpc \citep{moran:newman01} and last year with a more extensive experiment the Cepheid distance is now $7.5~\pm~0.15~\pm~0.17$ Mpc \citep{moran:macri06}.  There is now really quite good agreement between the Cepheid distance and the maser geometric distance.  The Cepheid distances given above are based on the distance to the Large Magellanic Cloud of 50 kpc, so one way of eliminating this small discrepancy between the maser and Cepheid distances would be to bring the Large Magellanic Cloud closer to us from 50 kpc by the ratio of 7.2/7.5.

\question{Unknown} How can you rule out that the actual disk is thicker than where the masers are occurring?

\answer{Moran} It is quite possible that the actual disk is thicker than the maser layer indicates.  My colleague Ramesh Narayan used to ask me this question frequently. His idea was that the masers are just a thin skin on a thicker accretion disk.  This would make the
accretion rate much higher than indicated by the thickness of the maser layer. I would like to
appeal to Occam's razor that the simple explanation is that the maser trace the true disk.  We
could possibly test the skin hypothesis when we get better data because you might think that the maser skin would be on the top on the redshifted side and on the bottom on the blueshifted side
so when you try to fit a model you would see a discontinuity. But this step would be very small.

\question{Kulkarni} How long do individual maser features last?

\answer{Moran} We don't know how long they last. Some have a lifetime of only a few years, but
some high velocity features have been there since we first saw them in 1992.  The systemic masers don't last much more than ten years.  It may be a rather turbulent medium and these blobs that produce the masers may come and go.

\question{Cordes} What's the gain of the masers that amplify the AGN?

\answer{Moran} It is rather controversial as to whether the systemic masers  actually amplify the
AGN emission or not.  The brightness temperatures of the masers aren't known because we don't resolve them but they must be at least $10^{14}$ K and the AGN can't be brighter than $10^{12}$ K so the gain is probably at least 100, and probably much higher. The high velocity masers have no apparent background sources to
amplify so their gain must be considerably higher.

\textbf{Acknowledgement.} I thank Alan Bridle for doing his editorial job very well, and Liz Humphreys for discussions and help with the figures.

\end{document}